\documentclass[10pt]{article}
\usepackage{amsfonts}

\usepackage{a4}
\usepackage{graphicx}
\usepackage{multicol}

\newlength{\figheight}
\newlength{\figwidth}
\begin{document}


\title{\Large\bf Single-molecule spectroscopy near structured dielectrics}
\author{\large Carsten Henkel$^1$\thanks{email: 
{\tt carsten.henkel@quantum.physik.uni-potsdam.de}} 
\ and Vahid Sandoghdar$^2$}
\date{\small $^1$ Institut f\"ur Physik, Universit\"at Potsdam,\\
Am Neuen Palais 10, 14469~Potsdam, Germany\\[1mm]
$^2$ Fakult\"at f\"ur Physik, Universit\"at Konstanz, Fach M696,\\
78457 Konstanz, Germany \\
[2mm] 12 August 1998}
\maketitle

\vspace*{-7cm}
{\hspace*{\fill} to be published in \emph{Optics Communications}}
\vspace*{6cm}

{\centerline{\bf Abstract}}
\begin{quotation}
\begin{quote}\small
We present an analytical approach to the calculation of the linewidth and
lineshift of an atom or molecule in the near field of a structured
dielectric surface. For soft surface corrugations with amplitude
$\lambda/50$, we find variations of the linewidth in the ten percent
region. More strikingly, the shift of the molecular resonance can reach
several natural linewidths. We
demonstrate that the lateral resolution is of the order of the
molecule-surface distance. We give a semiquantitative
explanation of the outcome of our calculations that is based on simple 
intuitive models. \\[5mm]
07.79.Fc -- Near-field scanning optical microscopes\newline
32.70.Jz -- Line shapes, widths, and shifts\newline
61.16.Ch -- Surface structure\newline
78.66.-w -- Optical properties of specific surfaces and microparticles
\end{quote}
\end{quotation}


\medskip\


\begin{multicols}{2}

\section{Introduction}

It is well accepted that the natural linewidth of the excited state of an
atom, as well as the exact value of its energy levels, are greatly
influenced by quantum fluctuations. When an atom is confined in a geometry
the existence of the boundary conditions for the electromagnetic field
results in modifications of the atomic radiative properties. There have been
many theoretical works on the calculation and interpretation of these
effects using quantum electrodynamics \cite{Milonni}. Much of the physics
involved can be addressed, however, by replacing a two-level atom by a
classical dipole moment and treating its radiation in the presence of
boundaries \cite{Chance78}. Such an approach 
is quite successful in treating the 
modification of spontaneous emission due to the new environment. 
The energy level shifts of the
atomic states in the near field can also be described very well using this
model \cite{Hinds91}. One finds the well-known Lennard-Jones potential which
is proportional to $1/z^{3}$. In the far field, however, the Casimir-Polder
shift, as well as the exact numerical value of the oscillatory resonant
coupling of the excited state can be obtained only from a fully quantum
electrodynamic treatment \cite{Hinds91,Casimir48}.

From the experimental side many groups have tried to study various aspects
of these phenomena in different systems. The first experimental evidence for
the modification of spontaneous emission was demonstrated in 1970 by
Drexhage \cite{Drexhage74}. Here a very thin layer of fluorescing ions were
separated from the underlying surface by a thin spacer layer, and the
emission lifetime was recorded for different spacings. This technique has
been used extensively ever since due to its simplicity and its very high
vertical spatial resolution \cite{Barnes98}. Direct experimental
verification of the energy level shifts of atoms in confined geometries was
also demonstrated successfully more recently by performing high resolution
spectroscopy \cite{Heinzen87,Sandoghdar92,Ducloy91b}. Following the
discovery of Surface Enhanced Raman Spectroscopy in the early eighties the
more complex case of a molecule in the vicinity of rough surfaces attracted
much attention. Several researchers have studied the emission properties of
an ensemble of dye molecules on rough surfaces and gratings \cite
{Knoll81,Metiu80,Metiu81,George87a}. Very recently there have been also some
efforts on the spectroscopy of atoms placed on a thin organic layer above a
rough surface \cite{Rubahn97}.

In this paper we treat the modification of the radiative properties of an
atom or a molecule placed very close to a surface with lateral optical
contrast. Our work is mainly motivated by the recent progress in the field
of Scanning Near-field Optical Microscopy (SNOM) which has opened the door
to optical microscopy and spectroscopy with lateral resolution beyond the
diffraction limit. In the most common SNOM configuration one arrives at this
high resolution by examining the sample in the near field of a
sub-wavelength metallic aperture. Indeed, single molecules on a surface have
been detected with this method, and it has been verified that the molecular
lifetime is modified by the presence of the aperture \cite{Ambrose94}. A
more elegant approach to SNOM uses the fluorescence of a single molecule as
a probe \cite{Kopelman93}. Here one can record the molecular emission
intensity or alternatively the molecular lifetime as a measure for the
interaction of the molecule with the sample surface. Girard and coworkers
have shown numerical calculations for the modification of the molecule's
lifetime as it is positioned above a sample with nanometric topographic
features \cite{Girard95,Rahmani97}. In the present paper we propose an
analytical approach for this problem based on a perturbative method from
scattering theory. Our approach is valid in the domain of soft surface
corrugations where a complex surface geometry can be Fourier decomposed in
terms of sinusoidal surface gratings whose corrugation amplitude is small
compared to the molecule-substrate distance. In addition to the modification
of spontaneous emission we also consider the modifications in the molecular
energy level shifts. The latter is particularly interesting in view of the
recent achievements in high resolution spectroscopy of single molecules \cite
{Basche}. In Konstanz we are currently pursuing experiments which aim at the
measurement of the energy level shifts of a single molecule in the vicinity
of a surface \cite{Sandoghdar98}. As we show in this paper one can take
advantage of the extremely high lateral resolution in this system to perform
a novel form of optical microscopy.

\section{Presentation of the model}

We are interested in the radiative properties of an atom or molecule (called
`molecule' in the following) at a position $\mathbf{r}$ in an inhomogeneous
environment. In this section we first outline the description of the
environment and then discuss the model taken for the molecule.

\subsection{Environment}

\begin{figure*}[tbp]
\centerline{
\resizebox{0.8\figwidth}{!}{
\includegraphics*[85,525][450,730]{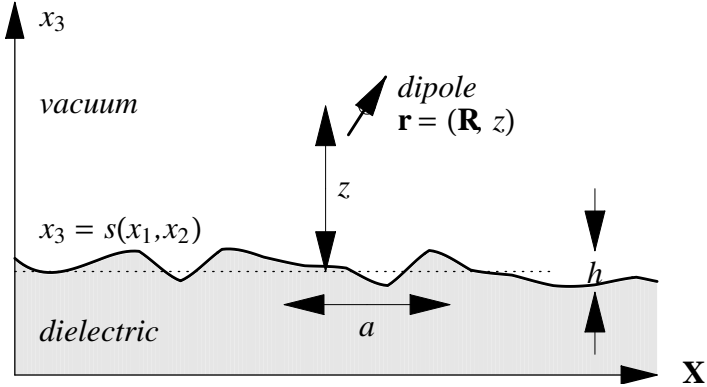}
}}
\caption{Geometry of the problem. The surface $x_{3}=s(x_{1},x_{2})$
separates the dielectric below (index $n$) from vacuum above. Three
different length scales are involved: the dipole's distance $z$ from the
mean surface, the vertical surface corrugation $h$ and the lateral
corrugation scale $a$. The transition wavelength $\protect\lambda $ is not
shown: typically, $z,\,h,\,a<\protect\lambda $ in near-field microscopy.}
\label{fig:geometry}
\end{figure*}

We consider the molecule to be placed in the vicinity of a solid substrate
at a distance ranging from a few nanometers to a few optical wavelengths
(see Fig.\ref{fig:geometry}). 
These distances are large compared to the dipole's dimensions, ensuring its
purely electromagnetic interactions with the surface. Moreover, it is
appropriate to describe the solid by a local dielectric function $%
\varepsilon (\mathbf{x};\omega )$, allowing the description of the
electromagnetic field phenomenologically by Maxwell equations.

The substrate surface is given by the equation $x_{3}=s(x_{1},x_{2})$ where $%
\mathbf{x}=(x_{1},x_{2},x_{3})$ are cartesian coordinates. The surface
corrugation is characterized by two length scales: the typical height $h$ of
the vertical corrugation and its lateral scale of variation $a$. In the case
of a surface relief grating $h$ would be the grating amplitude and $a$ its
period. In the region below the surface, $x_{3}<s(x_{1},x_{2})$, the
dielectric function is equal to the squared index of refraction $n^{2}$ that
is assumed to be real. The dipole is located in vacuum at the position $%
\mathbf{r}=(x,y,z)$. We explicitly retain its lateral coordinates $x,y$
since we are interested in the lateral resolution obtained in the variations
of the linewidth and lineshift. In order to simplify the formulas, we shall
use the notations $\mathbf{X}=(x_{1},x_{2})$ and $\mathbf{R}=(x,y)$ for the
lateral coordinates.

\subsection{Linewidth and frequency shift}

To begin with, let us focus on a system with two energy-levels, a ground
state $E_{\mathrm{g}}$ and an excited state $E_{\mathrm{e}}$ connected by an
electric dipole transition. We assume that the transition dipole is oriented
parallel to the $x_{j}$-axis ($j = 1,2,3$, linear polarization) 
and write $\mathcal{D}%
_{j} $ for the dipole matrix element. When this atom interacts with the
electromagnetic field its energy levels get shifted by amounts $\delta E_{%
\mathrm{g,e}}(\mathbf{r})$, and the excited state acquires a finite lifetime 
$1/\Gamma _{j}(\mathbf{r})$ where $\Gamma _{j}(\mathbf{r})$ is the
spontaneous emission rate. Both linewidth and level shifts may be calculated
in second-order perturbation theory. One obtains
the spontaneous emission rate \cite{Heitler} 
\begin{equation}
\Gamma _{j}(\mathbf{r})=\frac{\mathcal{D}_{j}^{2}}{\hbar ^{2}}%
\int\limits_{-\infty }^{\infty }\!\mathrm{d}\tau \,\langle \mathrm{vac}%
|E_{j}^{(+)}(\mathbf{r},t+\tau )E_{j}^{(-)}(\mathbf{r},t)|\mathrm{vac}%
\rangle \mathrm{e}^{\mathrm{i}\omega _{\mathrm{eg}}\tau }  \label{eq:gamma-e}
\end{equation}
where $\mathbf{E}^{(\pm )}(\mathbf{r},t)$ are the positive and negative
frequency parts of the electric field operator at the atom's position $%
\mathbf{r}$, and $\omega _{\mathrm{eg}}=(E_{\mathrm{e}}-E_{\mathrm{g}%
})/\hbar $ is the atomic transition frequency. At this point one often
proceeds to a mode expansion of the electric field operator, and the
linewidth~(\ref{eq:gamma-e}) connects to squared mode function amplitudes.
We take here a different route, following the response theory developed by
Agarwal \cite{Agarwal75a}, and Wiley and Sipe \cite{Sipe84}. More
specifically, we invoke the the fluctuation--dissipation--theorem to connect
the linewidth~$\Gamma _{j}$ to the classical Green function $G_{ij}(\mathbf{x%
},\mathbf{r};\omega )$. This Green function describes the electric field $%
\mathbf{E}_{\mathrm{dip}}(\mathbf{x})\mathrm{e}^{-\mathrm{i}\omega t}+%
\mbox{c.c.}$ (the `dipole field') created by an oscillating point dipole $%
\mathbf{d}\mathrm{e}^{-\mathrm{i}\omega t}+\mbox{c.c.}$ located at $\mathbf{r%
}$: 
\begin{equation}
{E}_{\mathrm{dip},i}(\mathbf{x})=\sum_{j}G_{ij}(\mathbf{x},\mathbf{r};\omega
)d_{j} .  \label{eq:def-Green}
\end{equation}
The fluctuation--dissipation--theorem allows one to express the linewidth~(%
\ref{eq:gamma-e}) in the following form \cite{Agarwal75a,Sipe84} 
\begin{equation}
\Gamma _{j}(\mathbf{r})=\frac{2\mathcal{D}_{j}^{2}}{\hbar }\mathop{\rm Im}%
\,G_{jj}(\mathbf{r},\mathbf{r};\omega _{\mathrm{eg}}) .
\label{eq:gamma-quantum}
\end{equation}
In the vicinity of an interface the electric field radiated by the dipole
differs from that in free space: it contains, in addition to the well-known
dipole field \cite{Jackson}, the field reflected from the surface. We write
this field in terms of a Green function $G_{ij}^{\mathrm{r}}(\mathbf{x},%
\mathbf{r};\omega _{0})$. Upon insertion into Eq.(\ref{eq:gamma-quantum}),
we find the environment-induced modification $\delta \Gamma _{j}(\mathbf{r})$
of the linewidth, that now depends on the atom's position relative to its
inhomogeneous environment. We stress that in the present approach, the
linewidth is linearly related to the electric field radiated by the dipole,
and it is not necessary to compute squared field mode amplitudes which is a
more difficult task in a complex geometry.

Let us now turn to the shift of the atomic resonance frequency $\delta
\omega _{\mathrm{eg},j}(\mathbf{r})=(\delta E_{\mathrm{e}}(\mathbf{r}%
)-\delta E_{\mathrm{g}}(\mathbf{r}))/\hbar $ in the presence of an
interface. By a calculation similar to the one for the linewidth, Fermi's
Golden Rule yields the following result (obtained from Eq.(2.9) of Ref.\cite
{Sipe84}): 
\end{multicols}
\begin{equation}
\delta \omega _{\mathrm{eg},j}=-\frac{\mathcal{D}_{j}^{2}}{\hbar }%
\mathop{\rm Re}\,G_{jj}(\mathbf{r},\mathbf{r};\omega _{\mathrm{eg}})+\frac{%
4\,\mathcal{D}_{j}^{2}}{\hbar }\int\limits_{0}^{\infty }\!\frac{\mathrm{d}%
\omega }{2\pi }\frac{\mathop{\rm Im}\,G_{jj}(\mathbf{r},\mathbf{r};\omega )}{%
\omega _{\mathrm{eg}}+\omega } .  \label{eq:domega-quantum-2-lev}
\end{equation}
The first term has a similar form as Eq.(\ref{eq:gamma-quantum}) but
involves the real part of the Green function.

\begin{figure*}[tbp]
\centerline{
\resizebox{0.5\figwidth}{!}{
\includegraphics*[90,620][260,770]{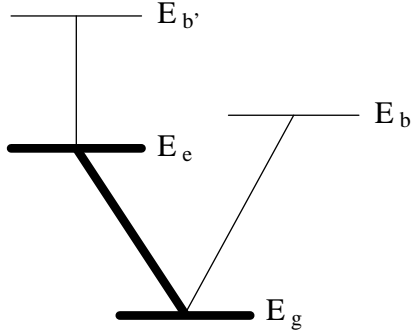}
}}
\caption[fig:atom]{Sketch of a multilevel atom with ground and first excited
state.}
\label{fig:atom}
\end{figure*}

For an atom with more than two states one has to take into account allowed
dipole transitions to other energy levels. Let us focus on the situation
depicted in Fig.\ref{fig:atom} 
where the excited level $E_{\mathrm{e}}$ is the first level above the ground
state $E_{\mathrm{g}}$. The linewidth $\Gamma _{j}$ is then still given by
the two-level expression~(\ref{eq:gamma-e}), and the lineshift contains an
additional contribution from higher-lying states $E_{\mathrm{b}}$ 
\begin{equation}
\delta \omega _{\mathrm{eg},j}^{\mathrm{other}}=-\frac{2}{\hbar }\sum_{%
\mathrm{b},\,E_{\mathrm{b}}>E_{\mathrm{e}}}\int\limits_{0}^{\infty }\!\frac{%
\mathrm{d}\omega }{2\pi }\mathop{\rm Im}\,G_{jj}(\mathbf{r},\mathbf{r}%
;\omega )\left( \frac{|\langle \mathrm{b}|d_{j}|\mathrm{e}\rangle |^{2}}{%
\omega _{\mathrm{be}}+\omega }-\frac{|\langle \mathrm{b}|d_{j}|\mathrm{g}%
\rangle |^{2}}{\omega _{\mathrm{bg}}+\omega }\right) .
\label{eq:domega-quantum-multi}
\end{equation}
\begin{multicols}{2}
\noindent
If the excited state decays to more than one lower-lying level, the decay
rate $\Gamma _{j}$ is a sum over several contributions, each one of
the form~(\ref{eq:gamma-quantum}).

It is instructive to compare the linewidth~(\ref{eq:gamma-quantum}) and the
frequency shift~(\ref{eq:domega-quantum-2-lev}) obtained from quantum theory
to the corresponding quantities for a classical harmonic oscillator. This
model of the Lorentz atom is widely used in the optics community \cite
{Chance78,Barnes98,Metiu84,Ford84}, and we can make contact with the work
done there. The Lorentz atom is a harmonic oscillator driven by the local
electric field, i.e., the external field plus the dipole field~(\ref
{eq:def-Green}). Since this field is proportional to the dipole moment
itself, it shifts the resonance frequency and leads to a finite damping rate.

For a linearly polarized dipole along the $x_{j}$-axis positioned in an
inhomogeneous environment, the shift $\delta \omega _{j}$ and the damping
rate $\Gamma _{j}$ are obtained from a straightforward calculation \cite
{Haroche92}. Normalizing to the free-space linewidth $\Gamma _{\infty }$,
one obtains for weak radiation damping (in SI units) 
\begin{equation}
\frac{\delta \omega _{j}(\mathbf{r})}{\Gamma _{\infty }}-\frac{\mathrm{i}}{2}%
\frac{\Gamma _{j}(\mathbf{r})}{\Gamma _{\infty }}=-\frac{3\pi \varepsilon
_{0}}{k_{0}^{3}}G_{jj}(\mathbf{r},\mathbf{r};\omega _{0})
\label{eq:result-Green}
\end{equation}
where $k_{0}=\omega _{0}/c$ is the vacuum wavenumber. As in the quantum
mechanical calculation, the imaginary part of the classical Green function
determines the linewidth. The fluorescence rate $\Gamma _{j}$ may
thus be computed classically \cite{Heitler}, and although we focus in the
following on the classical dipole model, our results for the linewidth
remain valid for a generic atom. As for the lineshift given by Eq.(\ref
{eq:result-Green}), the classical calculation only yields the first term of
the quantum-mechanical results~(\ref{eq:domega-quantum-2-lev}, \ref
{eq:domega-quantum-multi}), and the nonresonant frequency integrals of the
Green function are not accounted for. This implies a limited validity of the
classical model for frequency shift calculations. In this paper, however, we
are mainly interested in studying the variations in the radiative properties
of an atom as a function of its lateral position very close to a structured
surface. It is quite common that the atom-surface distances are much smaller
than the relevant atomic transition wavelengths. In this regime, the full
quantum-mechanical lineshift~(\ref{eq:domega-quantum-2-lev}, \ref
{eq:domega-quantum-multi}) approximately yields the electrostatic result 
\begin{equation}
\delta \omega _{\mathrm{eg},j}\approx -\frac{\langle e|d_{j}^{2}|e\rangle
-\langle g|d_{j}^{2}|g\rangle }{\hbar }\mathop{\rm Re}\,G_{jj}(\mathbf{r},%
\mathbf{r};\omega =0) .  \label{eq:domega-q-estatic}
\end{equation}
Note that this shift is again determined by a classical Green function, in
this case at zero frequency. The prefactor is different, however, from the
classical dipole and involves the difference in size of the electronic wave
function in the ground and excited states (a difference that cancels for a
two-level atom \cite{Barton74}). As far as the lateral resolution is
concerned, we may therefore treat the case of a classical dipole and use Eq.(%
\ref{eq:result-Green}) for simplicity. The exact magnitudes of the line
shifts and linewidths for realistic molecules can be then easily calculated
considering the above-mentioned discussion.


\section{Reflected field calculation}

\subsection{Outline}

Our task is now to compute the Green function above the substrate, i.e., the
reflected field created by an oscillating point dipole. The basic equations
are the macroscopic Maxwell equations, given the dielectric function $%
\varepsilon (\mathbf{x};\omega )$ and the external current $\mathbf{j}(%
\mathbf{x},t)=-\mathrm{i}\omega _{0}\mathbf{d}\delta (\mathbf{x}-\mathbf{r})%
\mathrm{e}^{-\mathrm{i}\omega _{0}t}+\mbox{c.c.}$, supplemented by the
boundary conditions for the field at the surface. We make the following 
\emph{ansatz} for the electric field in vacuum above the solid 
\begin{equation}
\mathbf{E}(\mathbf{x},t)=\left( \mathbf{E}^{\mathrm{fs}}(\mathbf{x})+\mathbf{%
E}^{\mathrm{r}}(\mathbf{x})\right) \mathrm{e}^{-\mathrm{i}\omega _{0}t}+%
\mbox{c.c.}  \label{eq:ansatz-Efield}
\end{equation}
where $\mathbf{E}^{\mathrm{fs}}(\mathbf{x})$ is the dipole field in free
space, and $\mathbf{E}^{\mathrm{r}}(\mathbf{x})$ is its environment-induced
modification (the reflected field). The latter is source-free above the
surface. The total field~(\ref{eq:ansatz-Efield}) is matched at the surface
of the solid to a `transmitted field' (source-free inside the solid). This
matching determines the reflected and transmitted fields in terms of the
free space dipole field.

In general the boundary conditions are complicated, and exact solutions are
only known for simple geometries. In order to proceed analytically, we
resort to an approximate solution for a `slightly corrugated surface',
i.e., a vertical corrugation
small compared to the separation of the molecule from the surface.
The precise validity of this approximation is discussed in Sec.\ref
{s:limitations}. The
boundary conditions may then be linearized around the mean value of the
surface function and solved to first order in the surface corrugation. We
thus end up with three terms for the linewidth 
\begin{equation}
\Gamma _{j}(\mathbf{r})=\Gamma _{\infty }+\delta \Gamma _{j}^{0}(z)+\delta
\Gamma _{j}^{1}(\mathbf{R},z)  \label{eq:Gamma-3-terms}
\end{equation}
The first term is the vacuum linewidth. The second comes from the reflection
at the flat surface (zeroth order field), and the third from the first-order
scattering off the surface corrugation. This last term contains information
about the lateral surface structure since it depends on the lateral
coordinate $\mathbf{R}=(x,y)$. A relation similar to~(\ref{eq:Gamma-3-terms}%
) also holds for the frequency shift $\delta \omega _{j}(\mathbf{r})$.

\subsection{Approximate solution}

In order to find the reflected field to first order, we use the so-called
`method of small perturbations' well-known in light scattering from rough
surfaces \cite{Maradudin75,Agarwal77,Nieto91}. With this method, one usually
computes the reflected field for an incident plane wave. We thus expand the
free space dipole field into Fourier components and work out the reflected
field for each component. This approach is similar to that of Rahman and
Maradudin \cite{Maradudin80a} and of George and co-workers \cite
{George87a,George87b}.

\subsubsection{Fourier expansions}

The free space dipole field is well known and may be found from the
following formula for the Green function (in SI units) \cite{Jackson,Nieto91}
\begin{equation}
G_{ij}^{\mathrm{fs}}( \mathbf{x}, \mathbf{r}; \omega ) = \left( \frac{
\partial^2 }{ \partial x_i \, \partial x_j } + \delta_{ij} k_0^2 \right) 
\frac{ \exp{\mathrm{i} k_0 | \mathbf{x} - \mathbf{r}| } }{ 4\pi\varepsilon_0
| \mathbf{x} - \mathbf{r}| }  \label{eq:Green-free-space}
\end{equation}
with the wavenumber $k_0 = \omega/c$. Using the notation $\mathbf{K} = (k_1,
k_2)$ for the lateral wave vector components and recalling the notations $%
\mathbf{x} = (\mathbf{X}, x_3)$, $\mathbf{r} = (\mathbf{R}, z)$ etc.\ for
the lateral and vertical coordinates, the Weyl expansion reads \cite{Nieto91}%
:
\begin{eqnarray}
\frac{ \exp{\mathrm{i} k_0 | \mathbf{x} - \mathbf{r}| } }{ 4\pi\varepsilon_0
| \mathbf{x} - \mathbf{r}| } 
& = & \int\!\frac{\mathrm{d}^2K}{ (2\pi)^2 } \frac{%
\mathrm{i}}{ 2\varepsilon_0\,k_3 } 
\label{eq:Weyl-expansion}\\
&& \times
\exp{\mathrm{i}[\mathbf{K} \cdot( \mathbf{%
X} - \mathbf{R} ) + k_3 |x_3 - z| ]} 
\nonumber
\end{eqnarray}
Here, the vertical wave vector component $k_3$ is defined by 
\begin{equation}
k_3 = \sqrt{k_0^2 - \mathbf{K}^2}  \label{eq:def-k3}
\end{equation}
and the square root is chosen such that $\mathrm{Re}\,k_3, \, \mathrm{Im}%
\,k_3 > 0$. It is important to note that the expansion~(\ref
{eq:Weyl-expansion}) contains both `far field' and `near field'
contributions, corresponding to lateral wave vectors with magnitude ${K}$
smaller and larger than the optical wavenumber $k_0$, respectively.

Combining Eqs.(\ref{eq:def-Green}, \ref{eq:Green-free-space}, \ref
{eq:Weyl-expansion}), we find the following Fourier expansion for the dipole
field in the region $x_3 < z$ below the dipole: 
\begin{equation}
\mathbf{E}^{\mathrm{fs}}( \mathbf{x} ) = \int\!\frac{\mathrm{d}^2K}{
(2\pi)^2 } \mathbf{E}^{\mathrm{fs}}( \mathbf{K} ) \exp{\mathrm{i}( \mathbf{K}
\cdot \mathbf{X} - k_3 x_3 ) }  \label{eq:Efs-expansion}
\end{equation}
where the `incident wave vector' is $\mathbf{k} = (\mathbf{K}, -k_3)$ and the
dipole field's Fourier components equal 
\begin{equation}
\mathbf{E}^{\mathrm{fs}}( \mathbf{K} ) = \frac{\mathrm{i} }{ 2 \varepsilon_0
\, k_3 } \left[ k_0^2 \mathbf{d} - \mathbf{k} (\mathbf{k} \cdot \mathbf{d}) %
\right] \exp{\mathrm{i}( - \mathbf{K}\cdot\mathbf{R} + k_3 z )}
\label{eq:Efs-Fourier}
\end{equation}
Since this vector is perpendicular to $\mathbf{k}$, it may be conveniently
expanded into two transverse polarization vectors $\mathbf{e}_\mu( \mathbf{K}
)$ labelled by the index $\mu = \mathrm{s}, \, \mathrm{p}$: 
\begin{eqnarray}
\mathbf{E}^{\mathrm{fs}}( \mathbf{K} ) 
& = & \frac{\mathrm{i} k_0^2 }{ 2
\varepsilon_0 \, k_3 } \sum_\mu \mathbf{e}_\mu( \mathbf{K} ) \left( \mathbf{e%
}_\mu( \mathbf{K} ) \cdot \mathbf{d} \right) 
\label{eq:Efs-Fourier-polar}\\
&& \times
\exp{\mathrm{i}( - \mathbf{K}%
\cdot\mathbf{R} + k_3 z )}  
\nonumber
\end{eqnarray}
To solve the boundary conditions, the expansion~(\ref{eq:Efs-expansion})
(and its counterpart~(\ref{eq:Er0-Fourier}) for the reflected field) is
assumed to be valid down to the surface $x_3 = s(\mathbf{X})$. This is
actually a hypothesis, the `Rayleigh hypothesis', as discussed by
Nieto-Vesperinas \cite{Nieto91}. Note that the present approach also relies
on the assumption that the dipole is located above the maximum surface
height, otherwise the absolute value $|s(\mathbf{X}) - z|$ in Eq.(\ref
{eq:Weyl-expansion}) must be retained which complicates the calculation.

As used by Agarwal \cite{Agarwal77} we apply the `extinction theorem' \cite
{BornWolf} and formulate the boundary conditions as integral equations
involving the field immediately above and below the surface. The integrand
is evaluated at the surface, and only zeroth and first order terms in the
profile function $s(\mathbf{R})$ are taken into account. The reflected field
thus contains zeroth and first order contributions that are discussed
separately in the following.

\subsubsection{Zeroth order: flat surface}

The \emph{zeroth order} result for the reflected field is obtained from the
reflection of each Fourier component from a flat surface and involves the
corresponding Fresnel coefficient $r_{\mu }(K)$. In the expansion 
\begin{equation}
\mathbf{E}^{\mathrm{r,0}}(\mathbf{x})=\int \!\frac{\mathrm{d}^{2}K}{(2\pi
)^{2}}\mathbf{E}^{\mathrm{r,0}}(\mathbf{K})\exp {\mathrm{i}(\mathbf{K}\cdot 
\mathbf{X}+k_{3}x_{3})}  \label{eq:Er0-Fourier}
\end{equation}
the Fourier coefficients are thus 
\begin{eqnarray}
\mathbf{E}^{\mathrm{r,0}}(\mathbf{K})
& = & \frac{\mathrm{i}k_{0}^{2}}{%
2\varepsilon _{0}\,k_{3}}\sum_{\mu }\mathbf{e}_{\mu }^{\mathrm{r}}(\mathbf{K}%
)r_{\mu }(K)\left( \mathbf{e}_{\mu }(\mathbf{K})\cdot \mathbf{d}\right) 
\label{eq:Er0}\\
&& \times\exp 
{\mathrm{i}(-\mathbf{K}\cdot \mathbf{R}+k_{3}z)}  
\nonumber
\end{eqnarray}
where the $\mathbf{e}_{\mu }^{\mathrm{r}}(\mathbf{K})$ are the unit
polarization vectors for the specularly reflected waves which are transverse
to the wave vector $(\mathbf{K},k_{3})$. From this result we read off the
reflected Green function for the flat surface and insert it into the general
formula~(\ref{eq:result-Green}), giving 
\end{multicols}
\begin{equation}
\frac{\delta \omega _{j}^{0}(z)}{\Gamma _{\infty }}-\frac{\mathrm{i}}{2}%
\frac{\delta \Gamma _{j}^{0}(z)}{\Gamma _{\infty }}=-\frac{3\pi \mathrm{i}}{2%
}\int \!\frac{\mathrm{d}^{2}K}{(2\pi )^{2}}\frac{\exp {(2\mathrm{i}k_{3}z)}}{%
k_{0}k_{3}}\sum_{\mu }{e}_{\mu ,j}^{\mathrm{r}}(\mathbf{K})r_{\mu }(K)\,{e}%
_{\mu ,j}(\mathbf{K}) .  \label{eq:flat}
\end{equation}
As expected, this result only depends on the dipole's distance $z$, and not
on its lateral coordinate $\mathbf{R}$. The integral over the azimuthal
angle of the two-dimensional wave vector $\mathbf{K}$ may be done
analytically, and one finds the familiar expressions for a dipole oriented
perpendicular ($\perp $) or parallel ($\Vert $) to the surface \cite
{Chance78}: 
\begin{eqnarray}
\frac{\delta \omega _{\perp }^{0}(z)}{\Gamma _{\infty }}-\frac{\mathrm{i}}{2}%
\frac{\delta \Gamma _{\perp }^{0}(z)}{\Gamma _{\infty }} &=&\frac{3\mathrm{i}%
}{4}\int\limits_{0}^{\infty }\!\mathrm{d}u\,\frac{u^{3}r_{\mathrm{p}}}{\sqrt{%
1-u^{2}}}\exp {(2\mathrm{i}k_{0}z\sqrt{1-u^{2}})}  \label{eq:gflat-perp} \\
\frac{\delta \omega _{\Vert }^{0}(z)}{\Gamma _{\infty }}-\frac{\mathrm{i}}{2}%
\frac{\delta \Gamma _{\Vert }^{0}(z)}{\Gamma _{\infty }} &=&\frac{3\mathrm{i}%
}{8}\int\limits_{0}^{\infty }\!\mathrm{d}u\,\frac{u}{\sqrt{1-u^{2}}}\left(
(1-u^{2})r_{\mathrm{p}}-r_{\mathrm{s}}\right) \exp {(2\mathrm{i}k_{0}z\sqrt{%
1-u^{2}})}  \label{eq:gflat-par}
\end{eqnarray}
\begin{multicols}{2}
\noindent
The integration variable is the reduced wave vector $u=K/k_{0}$. Note that
the integration range $0\leq u\leq n$ corresponds to field modes that are
plane waves in at least one half-space: the linewidth only depends on these
modes. For modes with larger wave vectors, $u>n$, the phase factor $\mathrm{e%
}^{2\mathrm{i}k_{3}z}$ and the reflection coefficients $r_{\mu }$ become
real and the integrands in Eqs.(\ref{eq:gflat-perp}, \ref{eq:gflat-par})
become purely imaginary. Hence these modes only appear in the lineshift.
This property does not hold any more when absorption in the dielectric 
is taken into
account \cite{Ford84,Haroche92} because the refraction index $n$ and hence
the reflection coefficients $r_{\mu }$ are complex for any wave vector.

\subsubsection{First order: lateral structure}

The \emph{first order} contribution to the reflected field is
Fourier-expanded as in Eq.(\ref{eq:Er0-Fourier}), and from the calculation
outlined above, one finds that the Fourier components equal \cite
{Agarwal77,Greffet88,vanLabeke93} 
\end{multicols}
\begin{equation}
\mathbf{E}^{\mathrm{r,1}}(\mathbf{K}^{\prime })=\mathrm{i}(n^{2}-1)\mathsf{L}%
(\mathbf{K}^{\prime })\cdot \int \!\frac{\mathrm{d}^{2}K}{(2\pi )^{2}}k_{0}s(%
\mathbf{K}^{\prime }-\mathbf{K})\,\mathbf{E}^{\mathrm{tr,0}}(\mathbf{K})\exp 
{\mathrm{i}(-\mathbf{K}\cdot \mathbf{R}+k_{3}z)} .  \label{eq:Er1-solution}
\end{equation}
We use the notation $\mathbf{K}$ for the wave vectors of the zeroth-order
field, as in Eq.(\ref{eq:Er0}), while $\mathbf{K}^{\prime }$ denotes the
wave vectors of the scattered field. In Eq.(\ref{eq:Er1-solution}), $s(%
\mathbf{K}^{\prime }-\mathbf{K})$ is the Fourier transform of the surface
profile, and $\mathbf{E}^{\mathrm{tr,0}}(\mathbf{K})$ is the Fourier
component of the field transmitted by the flat surface given by the Fresnel
coefficients $t_{\mu }(K)$. Finally, $\mathsf{L}(\mathbf{K})$ is a $3\times
3 $-matrix given by \cite{Agarwal77} 
\begin{equation}
L_{ij}(\mathbf{K})=\frac{k_{0}}{k_{3}+k_{3n}}\mathbb{P}_{ij}^{\Vert }+\frac{%
n^{2}K^{2}\mathbb{P}_{ij}^{\perp }-K_{i}K_{j}-(n^{2}k_{3}K_{i}\delta
_{3j}+k_{3n}K_{j}\delta _{3i})}{k_{0}(n^{2}k_{3}+k_{3n})}
\label{eq:L-operator}
\end{equation}
where $\mathbb{P}^{\Vert ,\perp }$ are projectors parallel and perpendicular to
the $xy$-plane (the mean surface) and $k_{3n}=\sqrt{n^{2}k_{0}^{2}-K^{2}}$
is the vertical component of the transmitted wave vector. It is understood
that the third component of the in-plane vector $\mathbf{K}$ vanishes.

From the first-order reflected field~(\ref{eq:Er1-solution}) we find the
following contribution to the Green function: 
\begin{eqnarray}
G_{ij}^{\mathrm{r,1}}(\mathbf{x},\mathbf{r};\omega ) &=&\mathrm{i}(n^{2}-1)%
\frac{k_{0}^{2}}{2\varepsilon _{0}}\int \!\frac{\mathrm{d}^{2}K^{\prime }}{%
(2\pi )^{2}}\sum_{k}{L}_{ik}(\mathbf{K}^{\prime })\int \!\frac{\mathrm{d}%
^{2}K}{(2\pi )^{2}}\frac{\mathrm{i}}{k_{3n}}k_{0}s(\mathbf{K}^{\prime }-%
\mathbf{K})  \nonumber \\
&&{}\times \sum_{\mu =\mathrm{s,\,p}}{e}_{\mu ,k}^{\mathrm{tr}}(\mathbf{K}%
)t_{\mu }(K){e}_{\mu ,j}(\mathbf{K})\exp {\mathrm{i}(\mathbf{K}^{\prime
}\cdot \mathbf{X}-\mathbf{K}\cdot \mathbf{R}+k_{3}^{\prime }x_{3}+k_{3}z)}
\label{eq:Gr1-solution}
\end{eqnarray}
\begin{multicols}{2}
\noindent
where the unit polarization vectors $\mathbf{e}_{\mu }^{\mathrm{tr}}(\mathbf{%
K})$ describe the field transmitted through the flat surface. Upon insertion
in our formula~(\ref{eq:result-Green}), one finds the first-order
contribution to the linewidth and the lineshift.

\subsection{Transfer function}

\label{s:transfer}

We observe in Eq.(\ref{eq:Gr1-solution}) that each wave vector $\mathbf{K}$
of the free-space dipole field is diffracted by the Fourier component $s(%
\mathbf{K}^{\prime }-\mathbf{K})$ of the surface profile in such a way that
the propagation from the dipole down to the surface and back again gives
rise to a phase factor $\exp {\mathrm{i}[(\mathbf{K}^{\prime }-\mathbf{K}%
)\cdot \mathbf{R} + (k^{\prime}_3 + k_3) z ] }$. This leads to a lateral
modulation of lineshift and -width at the `grating vector' $\mathbf{Q}=%
\mathbf{K}^{\prime }-\mathbf{K}$. It is expedient to choose this wave vector
as integration variable in~(\ref{eq:Gr1-solution}). One obtains the
following form for the first-order contribution to linewidth and -shift: 
\begin{equation}
\frac{\delta \omega _{j}^{1}(\mathbf{r})}{\Gamma _{\infty }}-\frac{\mathrm{i}%
}{2}\frac{\delta \Gamma _{j}^{1}(\mathbf{r})}{\Gamma _{\infty }}=\int \!%
\frac{\mathrm{d}^{2}Q}{(2\pi )^{2}}k_{0}s(\mathbf{Q})F_{j}(\mathbf{Q};z)\exp{%
\mathrm{i}(\mathbf{Q}\cdot \mathbf{R})}  \label{eq:transfer}
\end{equation}
where a dimensionless transfer function 
\begin{eqnarray}
F_{j}(\mathbf{Q};z) &=&\frac{3\pi (n^{2}-1)}{2}\int \!\frac{\mathrm{d}^{2}K}{%
(2\pi )^{2}}\frac{\exp {\mathrm{i}(k_{3}^{\prime }+k_{3})z}}{k_{0}k_{3n}} 
\label{eq:def-transfer} \\
&&\times \sum_{\mu =\mathrm{s,\,p}}\sum_{k}{L}_{jk}(\mathbf{K}^{\prime }){e}%
_{\mu ,k}^{\mathrm{tr}}(\mathbf{K})t_{\mu }(K)\,{e}_{\mu ,j}(\mathbf{K})
\nonumber
\end{eqnarray}
has been introduced. In this formula, it is understood that $\mathbf{K}%
^{\prime }=\mathbf{K}+\mathbf{Q}$ and $k_{3}^{\prime }$ is the corresponding
vertical wave vector component (eq.(\ref{eq:def-k3})). Eq.(\ref{eq:transfer})
shows that $F_{j}(\mathbf{Q})$ determines the relative contribution of the
profile's Fourier components $s(\mathbf{Q})$ to the linewidth and -shift. Of
particular interest is the width of this ``filter'' as a function
of the grating vector $\mathbf{Q}$ since it determines the lateral
resolution of the image.

\begin{figure*}[tbh]
\centerline{
\resizebox{!}{0.8\figheight}{%
\includegraphics*[85,570][265,765]{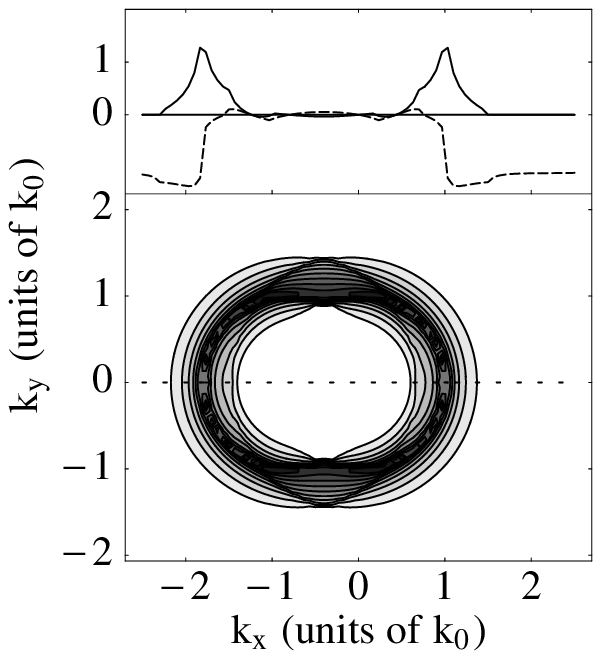}
}
\hspace*{1mm}
\resizebox{!}{0.8\figheight}{%
\includegraphics*[90,560][420,770]{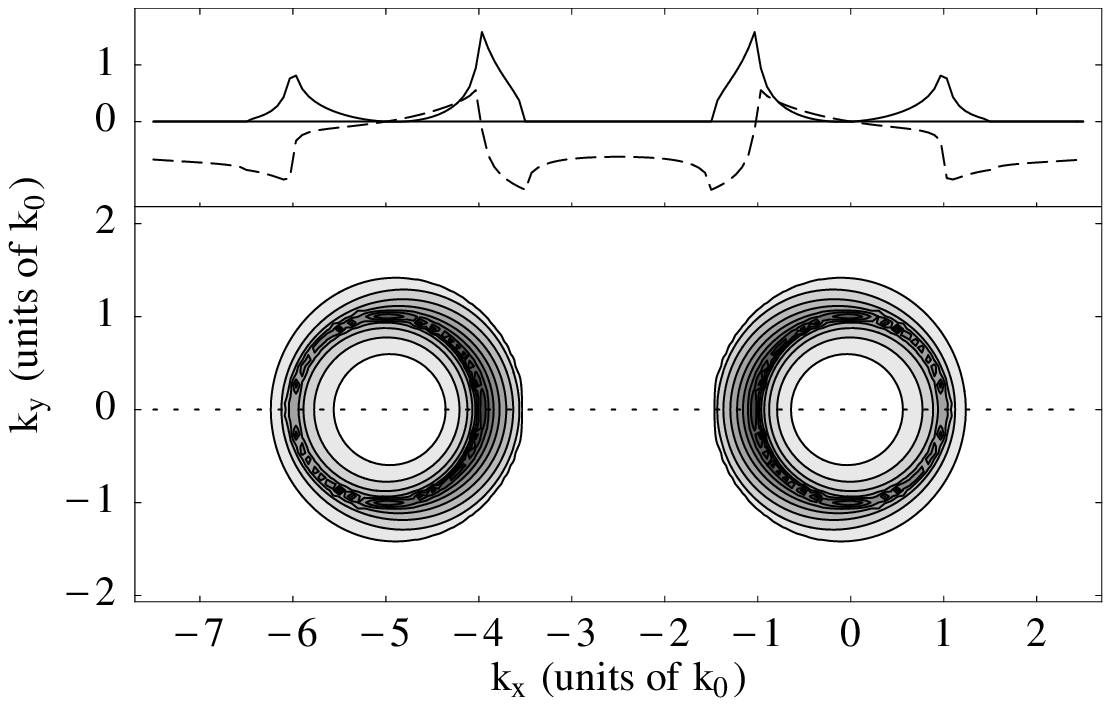}
}}
\caption[fig:integrand]{Integrand of the transfer function $F_{3}(\mathbf{Q}%
;z) $~(\ref{eq:def-transfer}) as a function of $\mathbf{K}$ in units of the
optical wave vector $k_{0}$,
for fixed grating vector ${\bf Q} \Vert {\bf e}_x$.
Left panel: grating period comparable to wavelength, $\mathbf{Q}%
=0.8\,k_{0}\,\mathbf{e}_{x}$. Right panel: subwavelength grating period, $%
\mathbf{Q}=5k_{0}\,\mathbf{e}_{x}$. 
The contours show the imaginary part, with dark
shading indicating large negative values. The inset shows a cut through the
dotted horizontal line (solid line: negative imaginary part, dashed line:
real part). \newline
The molecule's distance from the mean grating surface is $k_{0}z=0.3$. Its
dipole moment oscillates perpendicular to the mean grating surface (along
the $x_{3}$-axis). The substrate index is $n=1.5$. We plot the integrand
multiplied by $(2\protect\pi k_{0})^{2}$. }
\label{fig:integrand}
\end{figure*}

We show in Fig.\ref{fig:integrand} contour plots of the imaginary part of
the integrand in Eq.(\ref{eq:def-transfer}) for two different 
grating vectors $%
\mathbf{Q}$. It is apparent that the integrand does not have a simple
angular dependence in the plane of wave vectors $\mathbf{K}$. This implies
that in contrast to the flat surface case Eq.(\ref{eq:flat}), the angular
integral cannot be done analytically here, and the transfer function has to be
computed numerically. 
In Fig.\ref{fig:integrand} one or two circular structures appear to dominate
the integrand, depending on the magnitude of the grating vector $\mathbf{Q}$%
. To interpret these features, we come back to the Green function~(\ref
{eq:Gr1-solution}) that describes the first-order reflected field. For a
given Fourier component $\mathbf{K}$ of the incident dipole field and a
given grating vector $\mathbf{Q}=\mathbf{K}^{\prime }-\mathbf{K}$ of the
surface profile the scattered field amplitude is equal to the profile's
Fourier amplitude $s(\mathbf{Q})$, multiplied by the product of two factors.
The first is an electromagnetic scattering factor represented by the matrix $%
\mathsf{L}(\mathbf{K}^{\prime })$, the transmission coefficients $t_{\mu
}(K) $ and polarization vectors $\mathbf{e}_{\mu }(\mathbf{K})$ for the flat
surface. The second factor is the exponential $\exp {\mathrm{i}%
(k_{3}+k_{3}^{\prime })z}$ describing the vertical propagation of the field
from the dipole to the surface and back again. The magnitude of the second
factor depends on the character (propagating or evanescent) of the incident
and diffracted waves. In particular, the wave vectors $k_{3}$ and $%
k_{3}^{\prime }$ become imaginary for large $K$, $K^{\prime}$, and the
exponential is very small. This limits the relevant range of wave vectors
that contribute to the integral.

In the case of the linewidth, the limitation is even more strict. It is
determined by the imaginary part of the integrand (cf.\ Eq.(\ref
{eq:def-transfer})) and is given by two circular domains of incident
wave vectors $\mathbf{K}$ with $K<nk_{0}$ or $K^{\prime }<nk_{0}$ because it
is only in these domains that the exponential $\exp {\mathrm{i}%
(k_{3}+k_{3}^{\prime })z}$ and the electromagnetic scattering factor become
complex. These regions are clearly visible in the right panel of Fig.\ref
{fig:integrand}. The grating vector $\mathbf{Q}$ is here sufficiently large
to separate the two disks. The disks are merged in the left panel because
the grating vector is smaller. To summarize, the lateral variation of the
linewidth above a corrugated surface is dominated by two different
processes: propagating Fourier components of the dipole field are diffracted
into evanescent waves and interfere with the dipole field (the right
circular disk of Fig.\ref{fig:integrand}, centered at $\mathbf{K}=\mathbf{0}$%
) and conversely, evanescent Fourier components of the dipole field are
diffracted into propagating waves (the left disk, centered at $\mathbf{K}=-%
\mathbf{Q}$).

\subsection{Scanning modes}

Up to now we have determined the fluorescence spectrum of a molecule at a
constant height $z$ above the structured surface. It is also possible to
perform these calculations for the more common SNOM scheme of `constant-gap'
where the separation of the molecule from the underlying surface profile is
kept at a value $d$. In the notation of the present paper one measures the
quantity $\Gamma _{j}(\mathbf{R},d+s(\mathbf{R}))$ where $z=d+s(\mathbf{R})$
is the vertical coordinate of the molecule. When calculating this linewidth
from Eq.(\ref{eq:Gamma-3-terms}) one has to take into account that our
theory only describes surface corrugations $s(\mathbf{R})$ small compared to
the gap $d$. We thus find a linewidth 
\begin{equation}
\Gamma _{j}(\mathbf{R},d+s(\mathbf{R}))=\Gamma _{\infty }+\delta \Gamma
_{j}^{0}(d)+s(\mathbf{R})\frac{\partial \delta \Gamma _{j}^{0}(d)}{\partial z%
}+\delta \Gamma _{j}^{1}(\mathbf{R},d)  \label{eq:Gamma-const-gap}
\end{equation}
The lateral structure is contained in the last two terms, the first of which
corresponds to the derivative of the flat-surface linewidth. It turns out
that the transfer function for the constant-gap mode can be written in the
following form 
\begin{equation}
\tilde{F}_{j}(\mathbf{Q};d)=F_{j}(\mathbf{Q};d)-F_{j}(\mathbf{0};d)
\label{eq:transfer-const-gap}
\end{equation}
where in the second term the constant-height transfer function~(\ref
{eq:def-transfer}) is evaluated at zero wave vector. For simplicity, we focus
on the constant-height mode in the rest of this paper .


\section{Imaging a grating}

\begin{figure*}[tbp]
\centerline{
\resizebox{\figheight}{!}{%
\includegraphics*[95,580][380,770]{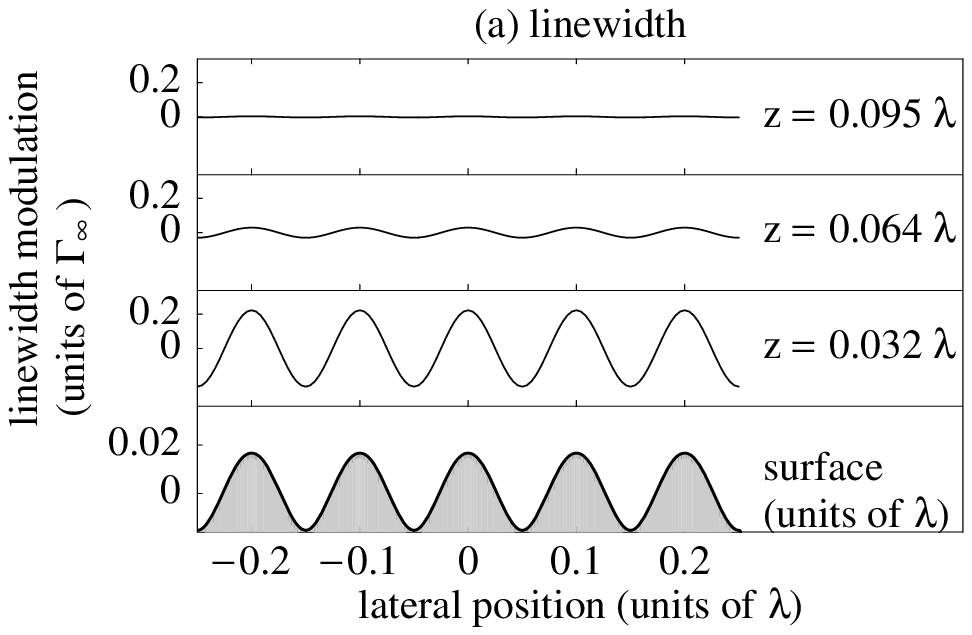}
}
\hspace*{2mm}
\resizebox{\figheight}{!}{%
\includegraphics*[95,580][380,770]{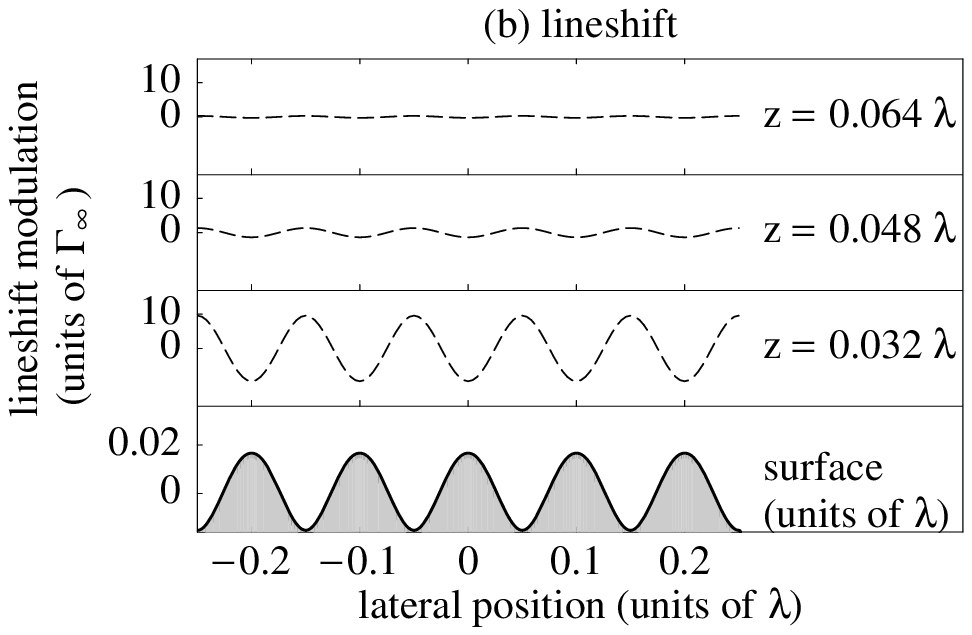}
}}
\caption[fig:grating]{The modulations in the linewidth $\protect\delta %
\Gamma ^{1}(x,z)$ (a) and lineshift $\protect\delta \protect\omega ^{1}(x,z)$
(b) of a dipole with vertical polarization at a fixed height above a
sinusoidal grating. The grating is made in a glass substrate ($n=1.5$) and
has period $a=\protect\lambda /10$ and amplitude $h=0.1\,\protect\lambda /(2%
\protect\pi )\approx 0.016\,\protect\lambda $. Note the difference in scale
between linewidth and -shift. }
\label{fig:grating}
\end{figure*}

We now examine linewidths and -shifts when the dipole is laterally scanned
at a constant height above a sinusoidal grating with surface profile $s(%
\mathbf{R})=s(x)=h\cos Qx$ (grooves parallel to the $y$-axis). This
simple geometry reveals the dependence of the radiative
properties on the four most important length scales 
(cf.\ fig.\ref{fig:geometry}): the
corrugation height $h$, the corrugation period $a$ (equal to $2\pi /Q$ for a
grating), the molecule's distance $z$ from the average surface and the
transition wavelength $\lambda =2\pi /k_{0}$. Finally, one also has to take
into account the dipole's orientation. Translational symmetry implies that
linewidth and -shift are independent of $y$, but they are sinusoidal as a
function of the lateral position $x$ since they depend linearly on the
surface profile (cf.\ Eq.(\ref{eq:transfer}) and Fig.\ref{fig:grating}). All
the relevant physics is thus encoded in their modulation amplitude \cite
{phaseshift}. Since this amplitude is simply proportional to the grating
height $h$, this length scale is already dealt with.

In Fig.\ref{fig:grating} the grating has subwavelength corrugation amplitude
($0.016\,\lambda $) and period ($0.1\,\lambda $), the substrate is a
dielectric with refractive index $n=1.5$ (glass) and negligible absorption,
while the dipole is polarized perpendicular to the grating surface ($z$%
-polarization). We observe that at an average distance from the grating of $%
0.032\,\lambda $, the linewidth modulation amounts to 20~\% of the natural
linewidth. A much larger modulation is observed in the lineshift which
amounts up to several natural linewidths. We would like to stress this
feature because frequency shifts have not been considered very much in the
optics community, perhaps because they are more difficult to measure. Our
calculations show that in near-field optics, the lineshift is much more
sensitive (on an absolute frequency scale) to the surface corrugation than
the linewidth. This relatively large effect is related to the large
frequency shift close to a flat surface, as discussed in subsection~\ref
{s:lineshift}. 
We discuss the dependence of linewidth and lineshift on the distance from
the grating and its period in the next two sections.


\subsection{The linewidth}

\begin{figure*}[tbp]
\centerline{
\resizebox{\figheight}{!}{%
\includegraphics*[95,565][410,775]{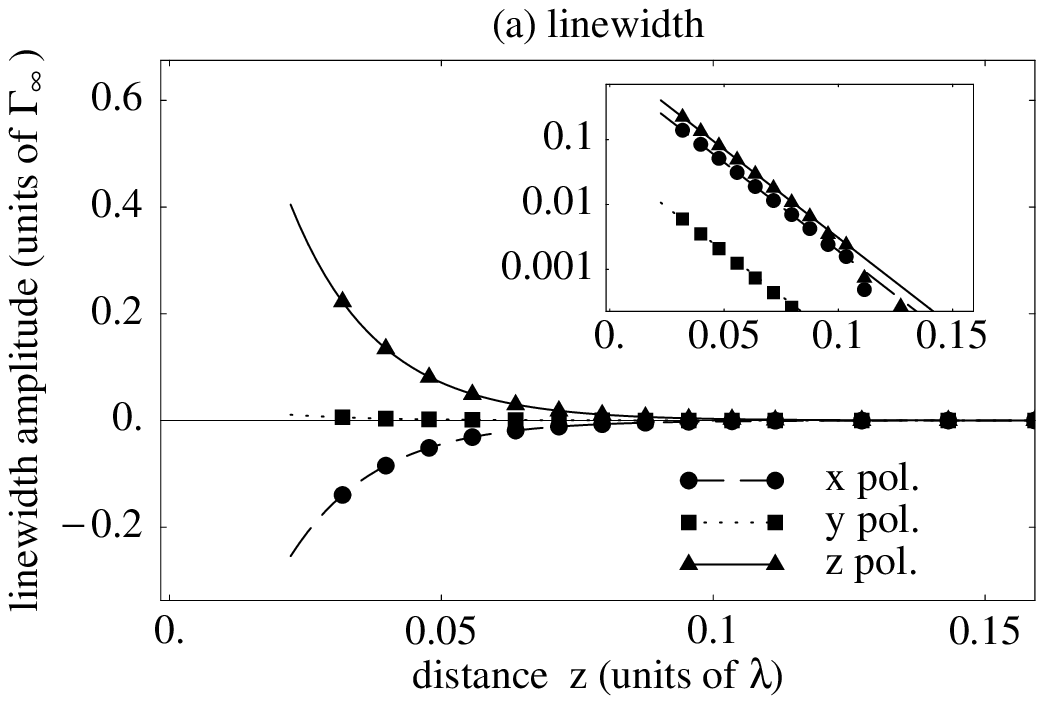}
}
\hspace*{2mm}
\resizebox{\figheight}{!}{%
\includegraphics*[95,565][410,775]{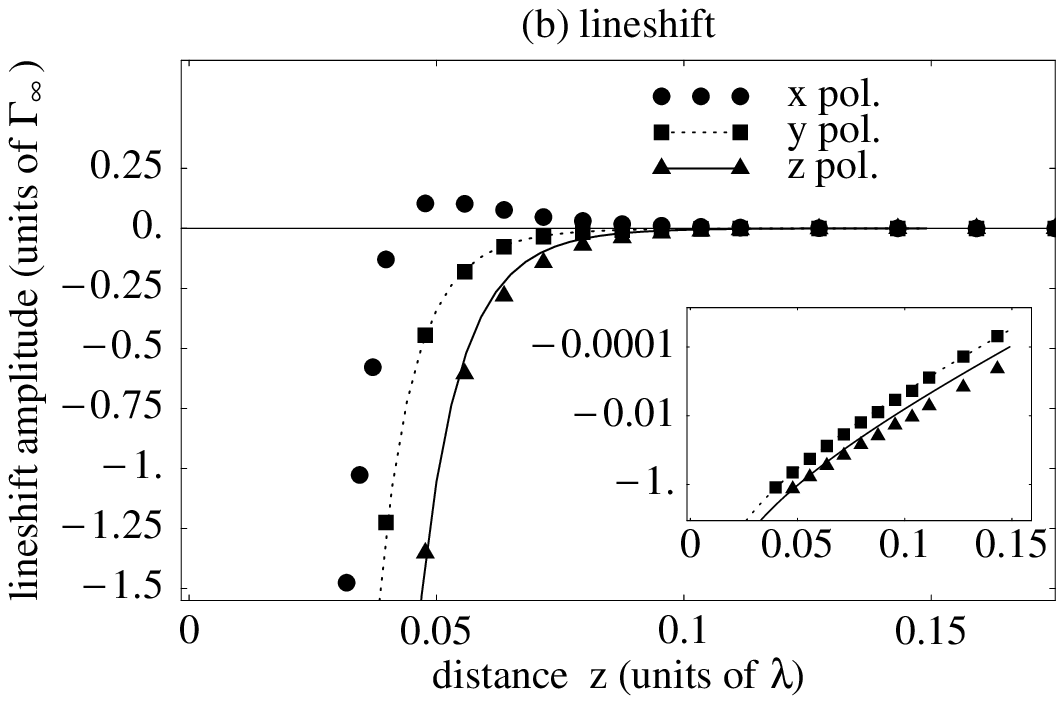}
}
}
\caption[fig:small-period]{The amplitude of the linewidth (a) and lineshift
(b) modulations vs.\ distance from a dielectric sinusoidal grating (index $%
n=1.5$) with period $a=\protect\lambda /10$ and amplitude $h=0.1\,\protect%
\lambda /(2\protect\pi )\approx 0.016\,\protect\lambda $. The three linear
polarizations are shown. The insets
give the same data on a logarithmic scale (the absolute value is taken).
In~(a), the lines are fits to simple exponentials $%
\mathrm{e}^{-Qz}$, the amplitude being the only free parameter. 
In~(b), the lines are fits to the model function~$K_{2}(Qz)/z^{2}$
introduced in eq.(\ref{eq:result-simple-domega}). For clearness,
the $x$-polarization is omitted from the inset. See text for more details. 
}
\label{fig:small-period}
\end{figure*}

In fig.\ref{fig:small-period}(a) we show the amplitude of the linewidth
modulation above a grating with subwavelength period, as a function of the
distance $z$. One observes that the three dipole orientations show different
behavior, and that the linewidth modulation decreases rapidly with
increasing distance from the grating. This decrease is quite well fitted
with an exponential law $\mathrm{e}^{-Qz}$, as shown in the inset. Such a
law is to be expected since as discussed at the end of 
section~\ref{s:transfer}%
, the linewidth is dominated by diffraction processes where the incoming
wave has a small parallel wave vector (see fig.\ref{fig:integrand}, right
panel). The diffracted wave then has a parallel component with wave vector $%
\mathbf{K}^{\prime }\approx \mathbf{Q}$ whose amplitude decays exponentially 
$\propto \mathrm{e}^{-Qz}$ if the grating vector $Q$ is much larger than $%
k_{0}$. This is the phenomenon which allows one to exploit the modifications
of the lifetime to image nanometric structures in the near field. Fig.\ref
{fig:small-period}(a) also shows deviations from a pure exponential decay of
the linewidth, we come back to these in eq.(\ref{eq:transfer-asymp}). 

\begin{figure*}[tbp]
\centerline{
\resizebox{\figheight}{!}{%
\includegraphics*[100,540][440,775]{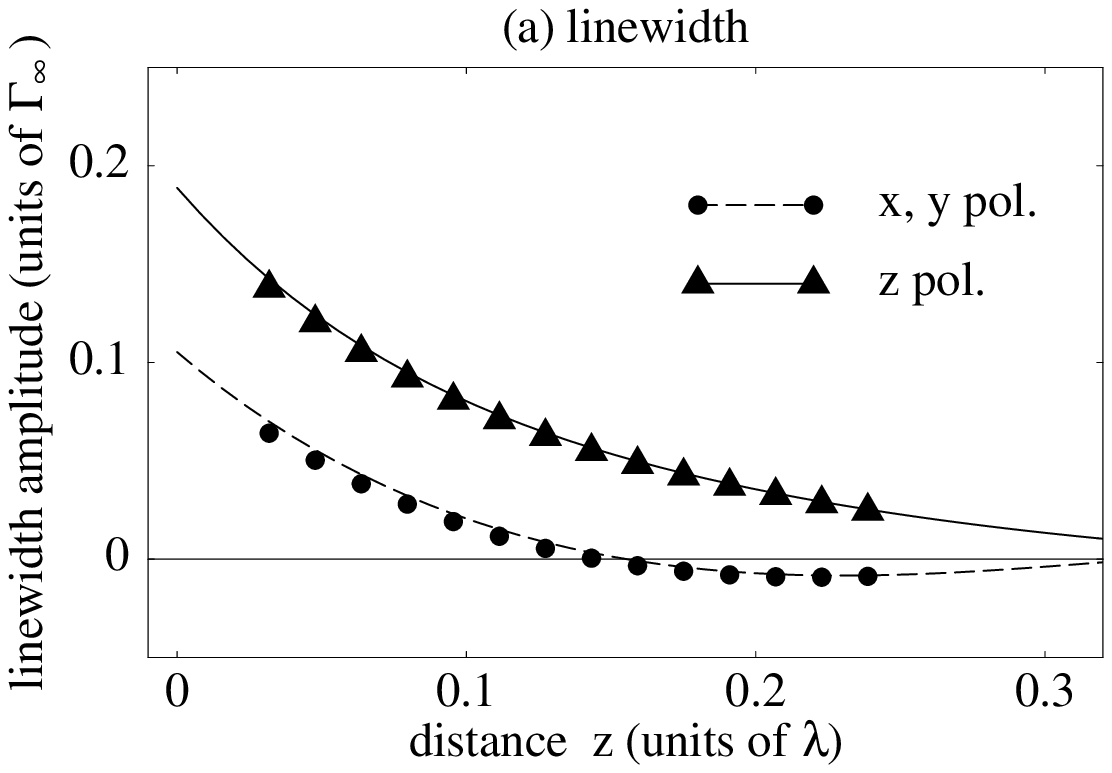}
}
\hspace*{2mm}
\resizebox{\figheight}{!}{%
\includegraphics*[100,540][440,775]{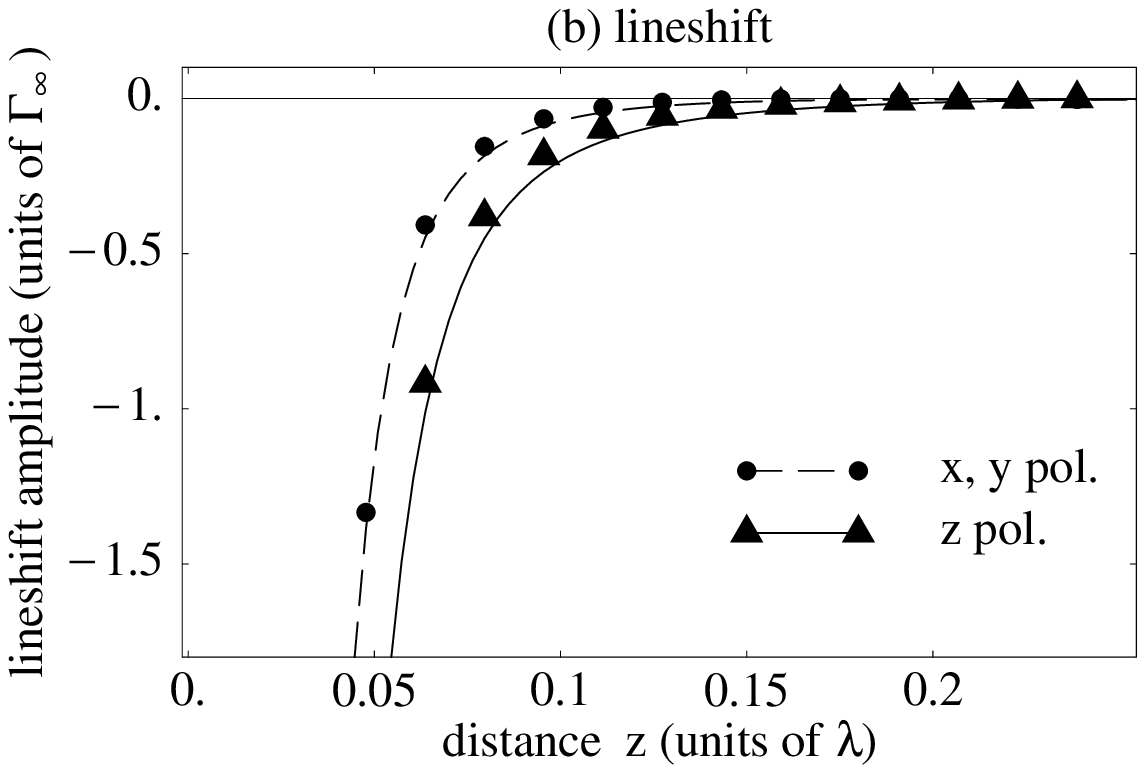}
}}
\caption[fig:large-period]{The amplitude of the linewidth modulation~(a) and
lineshift modulation~(b) vs.\ distance $z$ from a dielectric sinusoidal
grating (index $n=1.5$) with period $a=5\,\protect\lambda $ and amplitude $%
h=0.1\,\protect\lambda /(2\protect\pi )\approx 0.016\,\protect\lambda $.
Thin lines: simple model~(\ref{eq:model-derivative}) involving the
derivative of the flat surface result (see text for details). }
\label{fig:large-period}
\end{figure*}

In Fig.\ref{fig:large-period}(a) we plot the amplitude of the linewidth
modulation as a function of the average distance $z$ for a large grating
period. 
One observes that the modulation amplitude decreases more slowly than in fig.%
\ref{fig:small-period}(a) and even shows oscillations for distances of the
order of the wavelength. In the extreme case of grating period much larger
than the wavelength this behavior of the linewidth may be understood in a
simple manner. In this case one may write the linewidth modulation in the
form 
\begin{equation}
a\gg \lambda :\quad \delta \Gamma (x,z)\approx \delta \Gamma
^{0}(z-s(x))\approx \delta \Gamma ^{0}(z)-s(x)\frac{\partial \delta \Gamma
^{0}}{\partial z}  \label{eq:model-derivative}
\end{equation}
The linewidth modulation amplitude turns out to be proportional to the
derivative of the flat surface result. In Fig.\ref{fig:large-period}(a), the
result of this model is indicated by the thin lines which coincide quite
well with the full calculation (symbols) although the grating period is
taken to be only $5\,\lambda $. We note, however, that in this simple model
the two lateral polarizations $x$ and $y$ are always degenerate since they
both derive from the linewidth $\delta \Gamma _{\Vert }^{0}(z)$ of a dipole
polarized parallel to a flat surface.

\begin{figure*}[tbp]
\centerline{
\resizebox{\figheight}{!}{%
\includegraphics*[95,560][410,775]{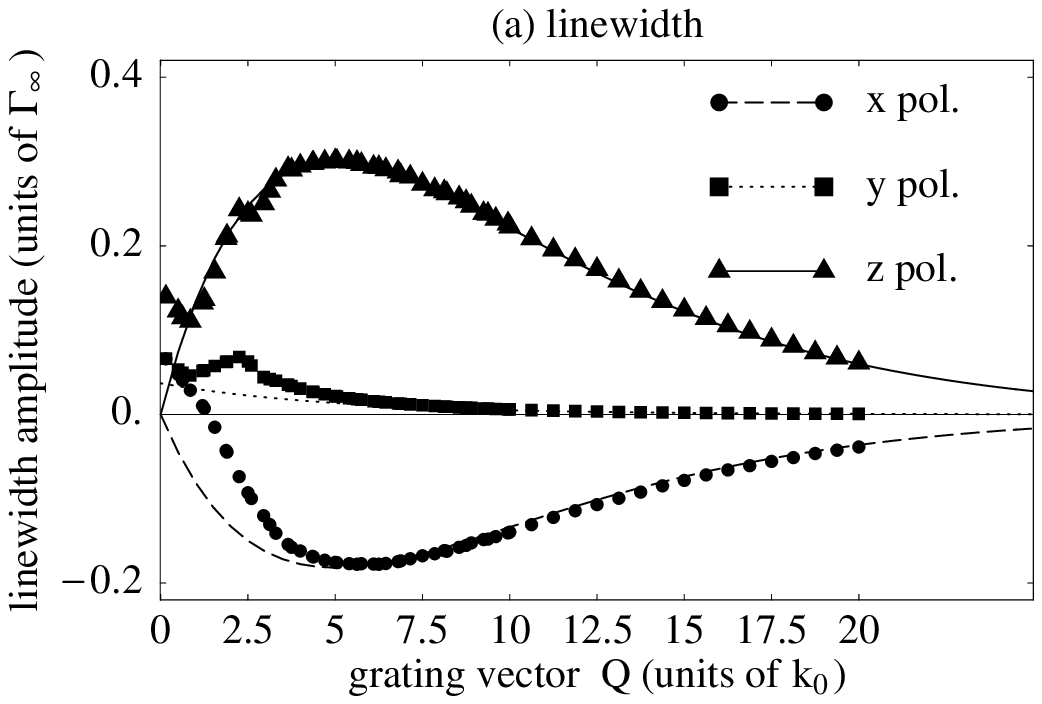}
}
\hspace*{2mm}
\resizebox{\figheight}{!}{%
\includegraphics*[95,560][410,775]{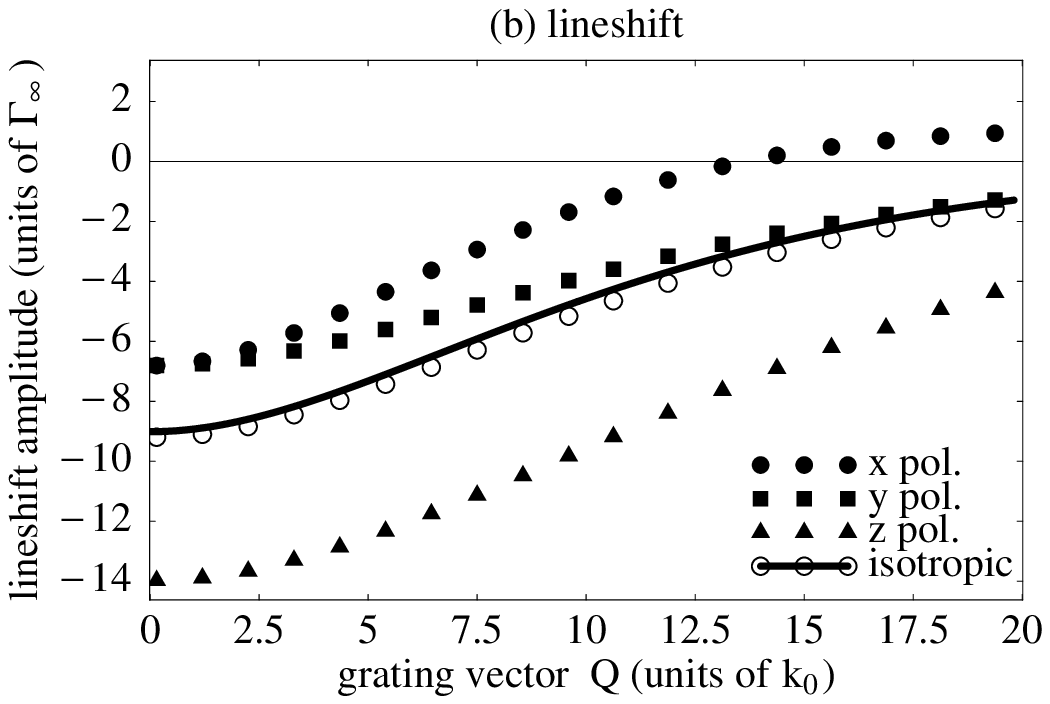}
}
}
\caption[fig:vs-Q]{ Amplitude of the linewidth modulation~(a) and lineshift
modulation~(b) vs.\ grating vector for a dipole at a distance of $z = 0.2\,%
\protect\lambda/(2\protect\pi) \approx 0.032\,\protect\lambda$ from the mean
surface. The three linear polarizations are shown. (a) The lines give the
linewidth modulation obtained from the asymptotic formula~(\ref
{eq:transfer-asymp}). (b) The open dots close to the thick line give the
lineshift for an isotropic dipole, and the thick line itself shows the
simple model of Eq.(\ref{eq:result-simple-domega}). The electrostatic value
of the coefficient $c_3$ is $( \Gamma_\infty / 8 k_0^3) (n^2-1)/(n^2+1)$. }
\label{fig:vs-Q}
\end{figure*}

Fig.\ref{fig:vs-Q}(a) shows the modulations of the linewidth as a function
of the grating vector $Q$ for a fixed height $z$. We observe that grating
periods larger than $\lambda$ ($Q \ll k_0$) yield a result similar to that
of a flat surface, $x$- and $y$-polarization being degenerate. As the period
decreases below the wavelength, this degeneracy is lifted, and we observe an
overall increase in the linewidth modulation with some steep features for $Q
\le 2nk_0$. 
For still smaller grating periods the linewidth modulation decreases again
in an approximately exponential manner.

It is possible to give an asymptotic expression for the transfer function~(\ref
{eq:def-transfer}) covering the regime of subwavelength corrugations which is
particularly interesting for applications in optical near-field microscopy.
This asymptotic expansion is motivated by Fig.\ref{fig:integrand} (right
panel) where we have seen that the integrand of the transfer function is
dominated by two circular regions of wave vectors. In the limit of $\mathbf{Q}
$ large compared to $k_{0}$ these regions are well separated and have
approximately circular symmetry. In both cases the angular integrations may
be performed analytically, and one arrives at formulae very similar to those
for a flat surface~(\ref{eq:gflat-perp}, \ref{eq:gflat-par}). For the three
linear polarizations one obtains: 
\end{multicols}
\begin{equation}
Q \gg k_0: \quad \mathop{\rm Im}\,F_{j}(\mathbf{Q};z) \approx \frac{3}{4}%
\frac{n^{2}-1}{n^{2}+1}\mathrm{e}^{-Qz}\left( \frac{Q}{k_{0}}
\right)^{\alpha_j} \mathop{\rm Im}\, f_{j}(z)  \label{eq:transfer-asymp}
\end{equation}
Here, the exponent is $\alpha_j = 1$ for the $j = x, z$ polarizations, and $%
\alpha_y = 0$ for $j = y$. If the distance $z$ is much smaller than the
wavelength, the (complex) dimensionless functions $f_{j}(z)$ ($j=x,y,z$)
that correct the pure exponential decay in eq.(\ref{eq:transfer-asymp}) are
given by \cite{corrections} 
\begin{eqnarray}
f_{x}(z) &=& 4 \int\limits_{0}^{n k_0 } \frac{ K\,\mathrm{d}K }{ k_0^2 } 
\mathrm{e}^{\mathrm{i}k_{3}z} \left( \frac{ k_0 }{k_{3}+k_{3n}} - \frac{ K^2 
}{ 2 k_{0} (n^{2}k_{3}+k_{3n}) } \right)  \label{eq:f1} \\
f_{y}(z) &=& - \left( n^2 + \mathrm{i} \right) \int\limits_{0}^{n k_0 } 
\frac{ K^3\,\mathrm{d}K }{ k_0^4 } \frac{ k_3 \, \mathrm{e}^{\mathrm{i}%
k_{3}z} }{ n^{2}k_{3}+k_{3n} } \\
f_{z}(z) &=& - 4 \mathrm{i} \int\limits_{0}^{n k_0 } \frac{ K^3\,\mathrm{d}K 
}{ k_0^4 } \frac{ n^{2} k_0\, \mathrm{e}^{\mathrm{i}k_{3}z} }{
n^{2}k_{3}+k_{3n} }  \label{eq:f3}
\end{eqnarray}
\begin{multicols}{2}
It can be seen from Fig.\ref{fig:vs-Q}(a) that these asymptotic formulae
give an excellent representation for the modulation of the linewidth above
subwavelength gratings with periods $a \le \lambda/5$. We may therefore use
eqs.(\ref{eq:transfer-asymp}--\ref{eq:f3}) 
to estimate the lateral resolution $\delta R$ for this type of
near-field microscopy: the exponential cutoff of high spatial frequencies in
Eq.(\ref{eq:transfer-asymp}) yields $\delta R \simeq z$, the distance from the
surface. Note also that the $y$-polarization gives both a smaller signal and
a slightly worse resolution compared to the other two polarizations, which
is due to the missing of the factor $Q/k_0$ in Eq.(\ref{eq:transfer-asymp}).
This is physically plausible because the electric field is then continuous
across the interface, leading to less scattering from the surface
corrugation.

\subsection{The lineshift}

\label{s:lineshift}

Apart from its larger modulation amplitude, the lineshift above a sinusoidal
grating shows a behavior not very different from that of the linewidth. In
fig.\ref{fig:small-period}(b) the grating period is subwavelength, and one
observes again a rapid decrease of the modulation amplitude with increasing
distance. Vector diffraction is relevant and leads to different results for
the three polarizations, the $x$-polarization in particular showing a sign
change (dots). At distances larger than $1/Q$, the lineshift shows an
exponential decrease similar to the linewidth, as shown by the solid and
dotted lines. These lines are fits to the model function $K_{2}(Qz)/z^{2}$
introduced in eq.(\ref{eq:result-simple-domega}) below.

The case of a grating period larger than the wavelength is shown in fig.\ref
{fig:large-period}(b). The $x$- and $y$-polarizations show identical
lineshifts as above a flat surface. The distance dependence is well
described by the simple model based on eq.(\ref{eq:model-derivative}),
involving the derivative of the flat-surface result. In particular, the
lineshift shows a $1/z^{4}$ power law if the dipole's distance is smaller
than about $\lambda /(2\pi )$, as expected from electrostatics.

Finally, fig.\ref{fig:vs-Q}(b) displays the amplitude of the lineshift
modulation if the grating period is varied. The $x,y$-polarizations show
smooth crossovers from large to small periods, and the modulation amplitude
globally decreases for very small periods. This latter feature can be
understood from a simple electrostatic calculation, as we discuss now.

We model the substrate as a continuous distribution of dipoles (with density 
$\rho_{\mathrm{dip}}$) with which the molecule interacts via a (scalar) $%
c_{6}/r^{6}$ law. The total frequency shift is obtained by integrating over
the half-space filled with these dipoles. For a flat substrate one obtains
the familiar power law $\delta \omega ^{0}(z)=-c_{3}/z^{3}$ with $c_{3}=\pi
\rho_{\mathrm{dip}}c_{6}/6$. If the substrate is corrugated, the surface
region gives an additional contribution to the frequency shift. To first
order in the surface profile, one obtains: 
\begin{equation}
z\ll \frac{\lambda }{2\pi }:\quad \delta \omega ^{1}(\mathbf{Q};z)\approx
-s(\mathbf{Q})\frac{3c_{3}}{2}\frac{Q^{2}}{z^{2}}K_{2}(Qz)
\label{eq:result-simple-domega}
\end{equation}
where $\delta \omega ^{1}(\mathbf{Q};z)$ is a Fourier transform with respect
to the lateral coordinates $\mathbf{R}$ and $K_{2}(Qz)$ is the modified
Bessel function of the second kind. As we show in fig.\ref{fig:vs-Q}(b),
this simple model (thick solid line) describes quite well the frequency
shift for an unpolarized dipole (averaged over the three linear
polarizations, shown by the open circles). For large periods the shift
becomes independent of $Q$ and tends to $-3c_{3}/z^{4}$, the derivative of
the flat-surface shift (this follows from the properties of the Bessel $K$%
-function) while for small periods we find an exponential suppression
similar to eq.(\ref{eq:transfer-asymp}): 
\begin{equation}
Qz\gg 1:\quad \delta \omega ^{1}(\mathbf{Q};z)\approx -s(\mathbf{Q})\frac{%
3c_{3}\sqrt{\pi }}{2\sqrt{2}}\frac{Q^{3/2}\mathrm{e}^{-Qz}}{z^{5/2}}
\label{eq:simple-domega-cutoff}
\end{equation}
In this model we may thus explain the suppression of high spatial
frequencies in the lineshift variations by the fact that the frequency shift
samples a patch of the surface whose radius is of order $z$, thus washing
out structures at lateral scales smaller than $a\leq z$. As a consequence,
we expect for lineshift images a lateral resolution of the order of $z$.

We finally note that similar to the linewidth~(\ref{eq:transfer-asymp}), the
lineshift~(\ref{eq:simple-domega-cutoff}) above a subwavelength grating does
not show a pure exponential decay with increasing distance $z$ (see also
inset of fig.\ref{fig:small-period}(b)).

\section{Imaging an arbitrary substrate}

\label{s:image}

\begin{figure*}[th]
\centerline{
\resizebox{\figwidth}{!}{%
\includegraphics*[85,475][395,755]{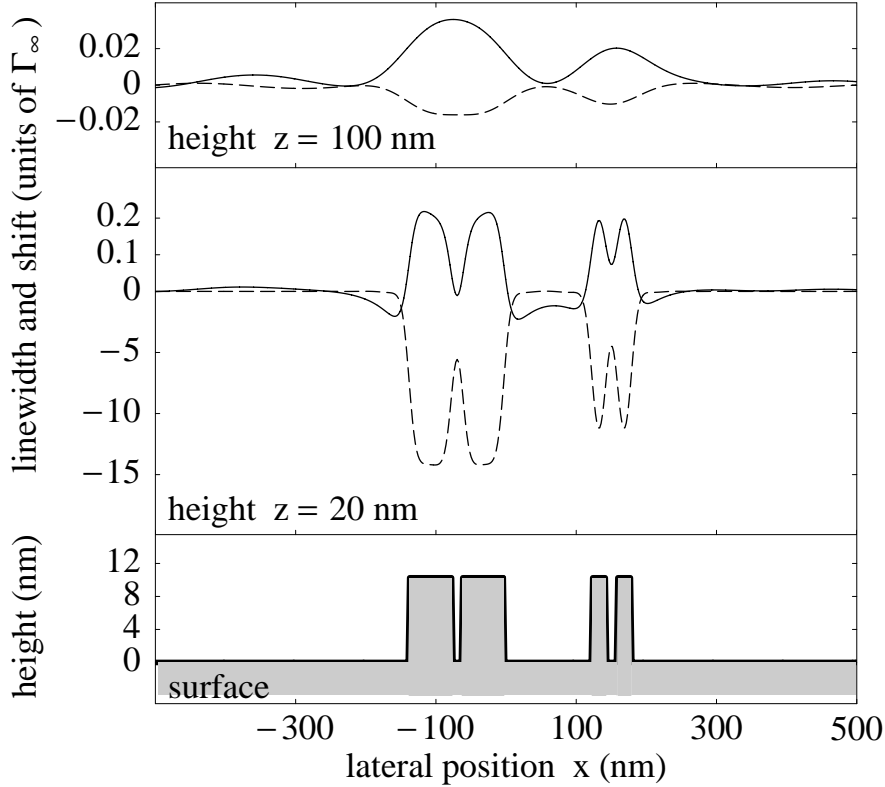}
}}
\caption[fig:images]{Linewidth and lineshift `images' of surface topography.
The surface profile is shown in the bottom panel (thick line, in nm). The
mid and top panels show linewidth (thin solid line) and lineshift (dashed
line) at two different constant heights above the mean surface, in units of
the free-space linewidth $\Gamma _{\infty }$. Only the laterally modulated
parts $\protect\delta \Gamma ^{1}(x,z)$ and $\protect\delta \protect\omega %
^{1}(x,z)$ are shown. Note the difference in scale for the lineshift at
height $z=20$~nm. Dielectric substrate with index $n=1.5$. Transition
wavelength $\protect\lambda =628$~nm. Vertical polarization. }
\label{fig:images}
\end{figure*}

As pointed out above, our theory is linear in the surface corrugation and
hence able to describe both sinusoidal and arbitrary profiles. An example of
a generic (two-dimensional) surface is shown in Fig.\ref{fig:images}. 
One observes a low lateral resolution and quite a weak signal at an average
height $z=100~\mathrm{nm}\approx \lambda /2\pi $ (top panel) with the
linewidth and -shift having comparable magnitudes. The situation changes
dramatically at closer distances ($z=20~\mathrm{nm}\approx 0.2\,\lambda
/2\pi $, middle panel) where subwavelength structures are well resolved.
Note that the linewidth gives a slightly poorer `image quality' than the
lineshift. This is due to the fact that the spectral response of the
lineshift behaves more smoothly as a function of wave vector than that of
the linewidth (compare figs.\ref{fig:vs-Q}(a) and~(b)). In other words, some
spatial frequencies are enhanced in the linewidth image, leading to a
distortion of the observed structures.

\begin{figure*}[th]
\centerline{
\resizebox{\figwidth}{!}{
\includegraphics*[85,470][400,760]{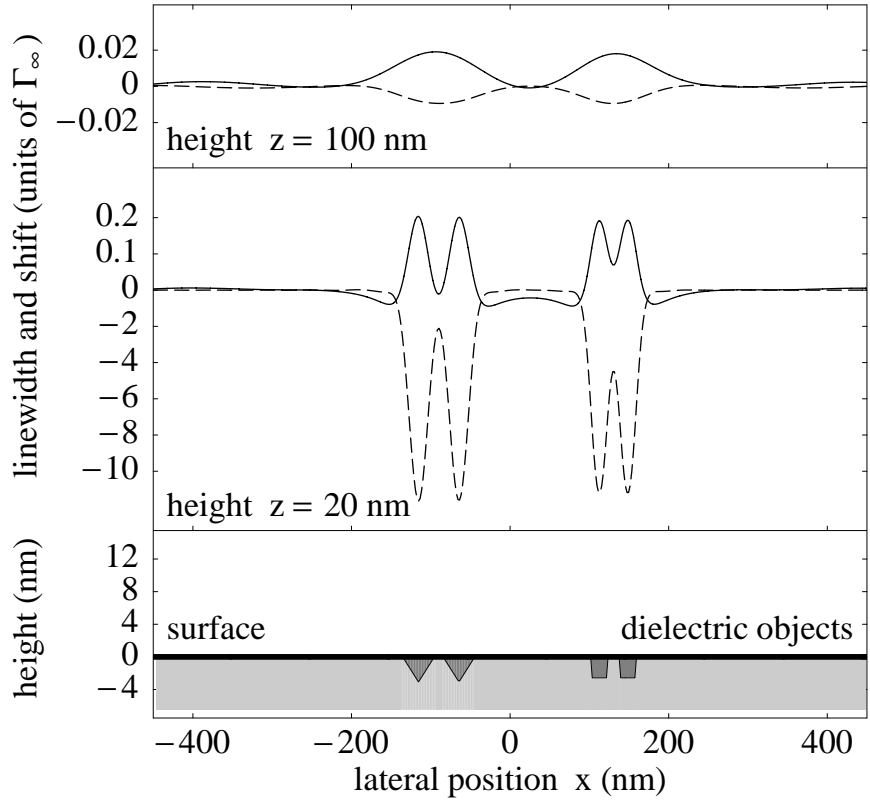}
}}
\caption[fig:opt-image]{Fluorescence image of a sample with optical
contrast. In the dark-shaded region, four objects with a larger index $n=2.5$
are buried in the flat substrate with $n=1.5$. Thin solid line: linewidth,
dashed line: lineshift. The other parameters are identical to fig.\ref
{fig:images}.}
\label{fig:opt-image}
\end{figure*}

One advantage of optical near-field microscopy over other scanning probe
techniques is its ability to yield information beyond the sample topography,
namely about its optical contrast. In fact, often samples with large
topographic features are undesirable in SNOM because they lead to the
coupling of the optical and topographic information 
\cite{Novotny97,Sandoghdar97}. It
is important to point out that our theory also applies to substrates with
purely optical contrast and no topography. For this we use the result of
Carminati and Greffet that in near-field optics variations of the dielectric
constant may be described by an `equivalent surface profile' $s_{\mathrm{eq}%
}(\mathbf{X})$ \cite{Greffet95c}. This quantity corresponds to the vertical
integral of the optical contrast: 
\begin{equation}
s_{\mathrm{eq}}(\mathbf{X})=\int\limits_{-\infty }^{\infty }\!\mathrm{d}%
x_{3}\,\left[ \varepsilon (\mathbf{X},x_{3})-\varepsilon _{\mathrm{flat}%
}(x_{3})\right]  \label{eq:def-equiv-profile}
\end{equation}
where $\varepsilon _{\mathrm{flat}}(x_{3})$ is the dielectric function of a
flat reference substrate. In fig.\ref{fig:opt-image} we show an example of
such a substrate where objects with larger indices are buried in a flat
substrate. 

\section{Limitations of the approximation}

\label{s:limitations}

The first-order calculation is crucially dependent on the linearized
boundary conditions at the `slightly corrugated surface'. More precisely,
this means that for all relevant Fourier components the following expansion
must remain sufficiently accurate: 
\begin{equation}
\exp {\ \mathrm{i}[(k_{3}+k_{3}^{\prime })s(\mathbf{X})] } \approx 1 + 
\mathrm{i}(k_{3}+k_{3}^{\prime })s(\mathbf{X})  \label{eq:hypothesis}
\end{equation}
where $k_{3}$ and $k_{3}^{\prime }$ are the wave vectors of the incident
dipole field and the diffracted field, respectively. For far-field
calculations, $k_{3}$ and $k_{3}^{\prime }$ are real and limited to the
optical wave vector $k_{0}$. Therefore, we obtain the condition $|s(\mathbf{X%
})|\leq h\ll \lambda /2\pi $, where $h$ characterizes the surface
corrugation. For calculations in the near field one has to include imaginary
values of $k_{3}$ and $k_{3}^{\prime }$. The \emph{relevant} wave vectors,
however, are limited in size because the finite distance $z$ leads to an
exponential damping; this is also apparent in Eq.(\ref{eq:def-transfer}). We
hence find the condition $h\ll z$. Finally, for a grating with a period $a$
well below $\lambda $ all diffraction orders are evanescent, and the
diffracted wave vectors are of the order of $k_{3}^{\prime }\simeq Q=2\pi /a$%
. In order to perform the linearization~(\ref{eq:hypothesis}) in this
regime, we have to impose the condition $h\ll a$. In summary, our method is
valid in the regime 
\begin{equation}
h\ll \min \{a,\,\lambda ,\,z\}  \label{eq:condition}
\end{equation}
Note that no restriction is made regarding the relative magnitude of the
three length scales on the right-hand-side. For a sample with optical
contrast it is shown in Ref.\cite{Greffet95c} that the perturbation method
is also subject to condition~(\ref{eq:condition}), but now for the
equivalent surface profile. In particular, this is the case if the index
inhomogeneities are confined to a narrow region around the interface, below
which the sample is homogeneous.

\section{Concluding remarks}

The theory presented here may be generalized to take absorption of the
substrate into account. In the classical picture this is simply done by
using a complex index of refraction. In the quantum mechanical picture
several schemes have been proposed \cite
{Barnett91,Yeung96,Knoell98a,Knoell98b} to quantize the electromagnetic
field in the presence of absorbing dielectrics, involving different models
for the dielectric medium. Using the theory of Scheel, Kn\"{o}ll, and Welsch 
\cite{Knoell98b}, it is easy to check that the fluctuation-dissipation
theorem still holds. The identification of the field correlation function in
Eq.(\ref{eq:gamma-e}) with the imaginary part of the classical Green
function in Eq.(\ref{eq:gamma-quantum}) hence carries over to the absorbing
substrate. From the viewpoint of quantum optics this substantiates the use
of the classical Lorentz oscillator to compute the fluorescence lifetime of
real molecules in arbitrary environments.

We now would like to remark on a few physical effects which take place
beyond the regime~(\ref{eq:condition}) and therefore, are not taken into 
account by our current treatment:

(i) If the grating corrugation is comparable to the wavelength $h\sim
\lambda $, one expects many diffraction orders to be populated. In this
regime the calculation of the reflected field has to be refined using a full
grating theory. At large distances from such a `deep grating' the physics
should be quite similar to standard far-field grating diffraction. At
smaller distances the molecule samples non-propagating diffraction orders,
and evanescent components of the incidence dipole field could lead to
qualitative changes of the linewidth. If the grating's `depth' exceeds
several wavelengths, one may expect the formation of a partial photonic band
gap, modifying the near field. Since for a complete band gap only evanescent
light modes are present, molecular fluorescence would be a very interesting
probe to study the electromagnetic field in such structures. In a future
paper we intend to consider a grating with a square profile for which an
exact diffraction theory is available \cite
{Wirgin69,Petit72,Botten81a,Sheng82}.

(ii) If the molecule is put into the selvedge region of the grating, i.e. $%
z\leq h$, it is nearly completely surrounded by the substrate. One then
expects that only the local environment plays a role, the molecule being
unable to sense the periodicity of the grating. Numerical calculations have
been done \cite{Girard95,Rahmani97,Novotny96} which show steep variations 
of the
lifetime and a strong polarization dependence. We plan to study the square
grating model alluded to above to get an analytical insight into this
situation.

In conclusion, we have calculated the modification of the fluorescence
spectrum of a molecule that is scanned above a slightly corrugated surface.
We have shown that the molecule's linewidth and lineshift are influenced by
both surface topography and optical contrast. The linewidth acquires
variations that amount up to 20~\%, while the lineshift varies over as much
as several natural linewidths. Furthermore, for an arbitrary surface profile
the lineshift shows a slightly better fidelity to the sample structure. We
have also presented simple models and formulae that allow us to obtain an
intuitive understanding of the our results for corrugations and distances
both below and above the molecular transition wavelength. Perhaps the most
important outcome of this paper is that the lateral resolution in the
`fluorescence images' we have obtained is of the order of the
molecule-surface distance. One would then expect to reach a molecular
resolution in a novel form of scanning optical microscopy if the probe
molecule could be brought nearly in contact with the sample. Recent
experimental progress in the field of single molecule detection,
spectroscopy and manipulation give a tantalizing hope for the realization of
this goal in the near future.

\section*{Acknowledgments}

It is a pleasure to thank J.-J.\ Greffet and M. Wilkens for useful
discussions and J. Mlynek for continuous support. We thank S. Scheel for
communicating a preprint of Ref.\cite{Knoell98b} prior to
publication and K. M\o lmer and J. Eisert for a careful reading of the
manuscript. We gratefully acknowledge the financial support of the Deutsche
Forschungsgemeinschaft (SFB~513 and He-2849/1-1).

\end{multicols}


\begin{thebibliography}{10}
\bibitem{phaseshift}
We have observed that for more complex polarizations, the modulations of
  linewidth and -shift are `phase-shifted' with respect to the grating profile.
  This is formally related to the fact that the Green
  tensor~(\ref{eq:Gr1-solution}) has nonzero off-diagonal elements.
\bibitem{corrections}
For distances not small compared to the wavelength, the integrands in
  eqs.~(\ref{eq:f1}--\ref{eq:f3}) acquire an additional $z$-dependence (not
  shown for brevity). The first corrections are of relative order $(n k_0
  z)^2$.
\end{thebibliography}
\end{document}